\documentclass[reprint, superscriptaddress, aps, showkeys, pra]{revtex4-2}

\usepackage[colorlinks=true, citecolor=blue, urlcolor=blue, linkcolor=magenta]{hyperref}
\usepackage{braket}
\usepackage{amsmath,amssymb,amsthm}
\usepackage{easybmat}
\usepackage[colorlinks=true,citecolor=blue,urlcolor=blue, linkcolor=magenta]{hyperref}
\usepackage[pdftex]{graphicx}
\usepackage{times,txfonts}
\usepackage{braket}
\usepackage{color}
\usepackage{natbib}
\usepackage{appendix}
\newcommand{\Edag}{\mathcal{E}^{\dagger}}
\newcommand{\fotoc}{\mathcal{F}(t)}

\begin{document}
    \title{Interferometric and Bipartite OTOC for Non-Markovian Open Quantum Spin-Chains and Lipkin-Meshkov-Glick Model}
\author{Baibhab Bose\textsuperscript{}}
\email{baibhab.1@iitj.ac.in}
\author{Devvrat Tiwari\textsuperscript{}}
\email{devvrat.1@iitj.ac.in}
\author{Subhashish Banerjee\textsuperscript{}}
\email{subhashish@iitj.ac.in}
\affiliation{Indian Institute of Technology Jodhpur-342030, India\textsuperscript{}}

\date{\today}

\begin{abstract}
    The information scrambling phenomena in an open quantum system modeled by Ising spin chains coupled to Lipkin-Meshkov-Glick (LMG) baths are observed via an interferometric method for obtaining out-of-time-ordered correlators ($\mathcal{F}-$OTOC). We also use an anisotropic bath connecting to a system of tilted field Ising spin chain in order to confirm that such situations are suitable for the emergence of ballistic spreading of information manifested in the light cones in the $\mathcal{F}-$OTOC profiles.  Bipartite OTOC is also calculated for a bipartite open system, and its behavior is compared with that of the $\mathcal{F}-$OTOC of a two-spin open system to get a picture of what these measures reveal about the nature of scrambling in different parameter regimes. Additionally, the presence of distinct phases in the LMG model motivated an independent analysis of its scrambling properties, where $\mathcal{F}-$OTOC diagnostics revealed that quantum chaos emerges exclusively in the symmetry-broken phase.
\end{abstract}
\maketitle

\section{Introduction}
The theory of quantum information scrambling has advanced the exploration of the mystery of quantum chaos~\cite{García-Mata:2023, Swingle_2024, mehta2004random, haake2001quantum, Toda_1988, Pandey_1981, pandey2019quantum, Zurek_1994, gutzwiller1991chaos}. It is known that chaotic quantum systems scramble local information fast to the whole system space~\cite{Swingle2018}. The considerable difference between chaotic and non-chaotic Hamiltonians is captured by different measures of quantum information scrambling. Two of the prominent witnesses of quantum chaos are OTOC (out-of-time ordered correlators)~\cite{García-Mata:2023, Swingle_2024, Hashimoto2017, Shenker2014, Standford_2015, Maldacena2016, Ueda_otoc, devvrat_OTOC}, and the bipartite OTOC~\cite{Zanardiopenbipotoc1, zanardibipotoc2, Bose_2024}, among others. These measures of information scrambling capture how local operators lose their characteristic locality with time under chaotic dynamics by employing an out-of-time ordered correlator of operators at different localities and times. The root of this concept lies in the growth of any local operator in the Heisenberg picture to the global Hilbert space and the consequent effect of the growth of commutativity of two initially non-commuting operators.
The black hole information paradox ~\cite{Maldacena2016, Hayden_2007}, and the problem of
quantum thermalization ~\cite{Srednicki_1994, Haake2, Hirsch_2022} have reignited curiosity in this area of study. One of the conventional methods for studying quantum chaos has been the random matrix theory ~\cite{mehta2004random, wigner, pandey2019quantum}. 

The OTOC is a four-point correlation function of local operators not following an order linear in time. It has been frequently applied in the theory of information scrambling, which facilitates deeper insights into quantum chaos~\cite{Hashimoto2017, Swingle_2024, Maldacena2016}. The OTOC was originally mentioned in the context of superconductivity ~\cite{Larkin1969QuasiclassicalMI} and made popular in~\cite{Shenker2014}. Analyzing the Lyapunov spectrum of the commutator growth in quantum systems that have a semiclassical limit is another crucial approach towards probing quantum chaos~\cite{Hirscheatomfield, Hirsch_2016, Tian2022}.  The study of several systems, including quantum field theories~\cite{Standford_2015}, has made substantial use of OTOCs. A few of these are random unitary models~\cite{David_2018}, spin chains ~\cite{Zhang_2019, Fortes_2020}, quantum phase transition models~\cite{otoc_floquet}, and quantum optical models ~\cite{Hirscheatomfield, devvrat_OTOC, Mahaveer}, among others ~\cite{García-Mata:2023, Swingle_2024, Swingle2018}. OTOC can take varied mathematical forms, like the regularized and the physical OTOC, wherein the fluctuation-dissipation theorem was generalized to the case of the OTOC~\cite{Ueda_otoc}. 
The quantity $\mathcal{F}$-OTOC, obtained from an interferometric scheme for calculating the OTOC, was presented in~\cite{SwingleInfoScram}. In \cite{Zhang_2019}, it was specifically generalized for the Gorini-Kossakowski-Sudarshan-Lindblad (GKSL) evolution~\cite{Lindblad1976, GKLSpaper}. 

Many qualitative features of the OTOC are insensitive to the specific operators chosen. A considerable simplification is thus provided by OTOCs averaged over random operators. The uniform averaging across pairs of random unitary operators, supported on both sides of the system's Hilbert space bipartition, can be carried out~\cite{zanardibipotoc2}. This averaged bipartite OTOC is operationally significant since it measures the operator entanglement of the dynamics and the information scrambling, given a general quantum channel~\cite{zanopen44, Zanardi2001opent}. Furthermore, any particular selection of unitary matrices to examine the information scrambling in the system is eliminated by Haar averaging over all of the unitary matrices in the specified Hilbert space. Recently, the equivalence between the chaotic behavior of OTOCs and the Wigner separability entropy (WSE) for a multipartite system was investigated~\cite{OTOC-RE}. This is the remnant of the OTOC-R\'enyi entropy theorem that deals with the averaged OTOC and its qualitative similarity with the second R\'enyi entropy.
To investigate chaos and information scrambling in many-body quantum systems with unitary operators, the averaged bipartite OTOC is a useful tool. The framework of ~\cite{zanardibipotoc2}, expanded to include open quantum systems~\cite{Zanardiopenbipotoc1}, i.e., the quantum systems impacted by their ambient environment, will be used in this paper.

The scrambling of quantum information in open quantum systems has been the subject of several recent studies~\cite{Zhang_2019,devvrat_OTOC,zanopen21, zanopen31, Bose_2024, infor_scram_non_Markov, Deffner_info_scram_otoc}. Thus, for example, an analog of the quantum regression theorem has been used to calculate the out-of-time-ordered correlations~\cite{QRT_OTOC_1, QRT_OTOC_2}.
The study of open quantum systems examines how quantum systems behave while accounting for the impact of their environment ~\cite {Weiss2011, Breuer2007, Banerjee2018}. This has made significant progress in recent years~\cite{Omkar2016, Javid_2018, Vacchini_2011, Tiwari_2023, tiwari2024strong}. The Gorini-Kossakowski-Sudarshan-Lindblad (GKSL) formalism is a commonly used approach to the dynamics of open quantum systems ~\cite{GKLSpaper, Lindblad1976}, where, using the Born-Markov and rotating wave approximations, a Markovian evolution is generated. With the advancement in theory and technology, consideration is being given to the effects of non-Markovian evolution~\cite{Hall_2014, Rivas_2014, RevModPhys.88.021002, CHRUSCINSKI20221, banerjeepetrucione, vega_alonso, Utagi2020, kading2025}.

Spin-chain models have served as a fundamental platform for exploring quantum chaos and information scrambling in quantum many-body systems~\cite{García-Mata:2023, Swingle2018}. In this context, the Lipkin-Meshkov-Glick (LMG)~\cite{LIPKIN1965188, Vidal_LMG, Hou2016} and the tilted field Ising models (TFIM) have been put to use in various scenarios~\cite{TiltedMfieldofarul, Bose_2024, bose2025, LMG_chaos1}. In this work, we consider two models: an Ising spin chain model surrounded by an LMG bath~\cite{Hou2016, Deffner_2017}, and the TFIM interacting with an anisotropic spin-chain bath at one end. Further, the quantum chaos of the LMG model, owing to its interesting properties, is also studied. The LMG model is a global interaction type model since its individual spins interact with all other spins, whereas the TFIM, due to nearest-neighbour interaction, is a local interaction type of model. 

The OTOC (interferometric) and the bipartite OTOC are diagnostics of quantum chaos for Hamiltonians made of local operators. It is well established that, for local Hamiltonians of chaotic nature, the OTOC $\mathcal{F}(t)$ decays while the bipartite OTOC $G(t)$ rises rapidly, signaling fast information scrambling. These quantities are quantum correlation functions of local operators located at different positions and evaluated at different times. Hence, the extent of decay of such correlations reflects the persistence of a local perturbation in the Ising spin, a quantum remnant of the classical phenomenon of sensitivity to initial conditions.  In this paper, we consider the Ising spin chain as such a system while different classes of baths are coupled to it. The idea is to observe the nature of these diagnostic tools for chaotic and non-chaotic models and how they are affected by the presence of a globally coupled LMG bath and a locally coupled spin chain bath. The LMG bath is a set of spins that globally interact with each other. We have analyzed the effect on OTOC and bipartite OTOC of Ising spin chains in an LMG bath and compared it with the effect of anisotropic spin chain baths.
Effects of system-bath interactions on information scrambling in these models are examined for arbitrary coupling strengths representing non-Markovian dynamics. Interferometric OTOC and bipartite OTOC are employed to examine information scrambling within these models. Additionally, the impact of phase transitions on information scrambling and chaotic behavior, particularly with respect to the LMG model, is analyzed.

The paper is arranged as follows. In Sec.~\ref {prelims}, we discuss the preliminaries, including $\mathcal{F}-$OTOC, bipartite OTOC, and the spin-chain and LMG models used in this paper. $\mathcal{F}-$OTOC and bipartite OTOC for the Ising spin-chain model interacting with the LMG bath are provided in Sec.~\ref{sec_III}, including a comparison between them. Section~\ref{Sec_IV} discusses $\mathcal{F}-$OTOC for TFIM interacting with an anisotropic spin chain. In Sec.~\ref{Sec_V}, $\mathcal{F}-$OTOC for the LMG model is explored, followed by the conclusion in Sec.~\ref{sec_conclusions}.

\section{Preliminaries}\label{prelims}
We now discuss briefly the two important measures of information scrambling, the $\mathcal{F}-$OTOC and the bipartite OTOC.
The interferometric OTOC is an experimental scheme for calculating the out-of-time-ordered correlator (OTOC), which is a four-point correlation function of local operators sequenced non-linearly in time. Initially correlated spatially separated local operators are seen to lose the correlation, as indicated by the rapid decay of OTOCs. This four-point correlation function has its root in the commutator square that corresponds to the sensitivity to the initial conditions in classical chaos. The same phenomenon of scrambling is observed when two initially non-commuting or independent local operators start commuting fast, implying de-localization of one of the local operators, as they grow through the composite Hilbert space. 
The bipartite OTOC is a sophisticated version of the commutator squared. The system is bipartitioned, and all the possible local unitary operators in those two bipartitions are Haar-averaged, thereby eliminating the specificity of the choice of the local unitary operators. Naturally, the gross nature of the bipartite OTOC remains the same as that of the commutator squared, i.e., it grows exponentially from zero for a fast scrambler, a chaotic Hamiltonian.

\subsection{$\mathcal{F}-$OTOC}
Consider a quantum many-body system of several spins, comprising a local Hamiltonian having nearest-neighbor interactions. Any initial perturbation in terms of local operations (e.g., spin-flip $\sigma_z$) spreads away to other sites. The Baker-Hausdorff-Campbell (BCH) expansion shows how an operator in the Heisenberg picture grows in the total Hilbert space while the initial operator $A(0)$ acts on a single site.
\begin{equation}
    A(t)= A + i t [H, A] + \frac{(i t)^2}{2!} [H, [H, A]] + \frac{(i t)^3}{3!} [H, [H, [H, A]]] + \dots\, .
\end{equation}
Any other local perturbing operator $B$ that acts on a different site can be used as the necessary counterpart to calculate the correlation between these two perturbations. The growth of this correlation reveals the nature of the system in terms of how a given perturbation is sustained over time and to what distance. Commutators are one of such correlation functions. Initially, when perturbations on different sites have not affected each other, they are said to commute with each other and are uncorrelated. Consequently, the growth of the commutator value relates to the growing correlation of the two operators, and the nature of that growth is a diagnostic for whether a system is chaotic or not.
The growth of the Hilbert-Schmidt norm of this commutator shows the degree to which the initial perturbation has affected the new perturbation at time $t$. 
This commutator growth that illustrates how the information of the perturbation scrambles through the system is~\cite{devvrat_OTOC}
\begin{equation}\label{commusquare}
    \mathcal{C}_{AB}(t)=\frac{1}{2}{\rm Tr} ( [A_t,B]^{\dagger}[A_t,B]\rho).
\end{equation}
Here, $A_t$ and $B$ are arbitrary Heisenberg local operators, $\langle \cdot \rangle = {\rm Tr}(\rho \cdot)$ is the average with respect
to the thermal density matrix of the system $\rho = e^{-\beta H} /Z$,
where $Z = {\rm Tr}(e^{-\beta H})$ and $\beta = 1/k_BT$. Further, $H$ is the Hamiltonian of the system.
Many features of a quantum many-body system, along with quantum chaos, are characterized by this measure of scrambling of quantum information.
Expanding this commutator structure into its four elements, we have
\begin{align}
    \mathcal{C}_{AB}(t) = \frac{1}{2}(\mathcal{D}_{AB}(t) + \mathcal{I}_{AB}(t) - 2 \, \Re \{ \mathcal{F}_{AB}(t) \}),
\end{align}
where $\mathcal{D}_{AB}(t) = \langle B^\dagger (A^\dagger A)_t B \rangle$, $\mathcal{I}_{AB}(t) = \langle A^\dagger{}_t (B^\dagger B) A_t \rangle$, $\mathcal{F}_{AB}(t) = \langle A^\dagger{}_t B^\dagger A_t B \rangle$, and $\Re\{*\}$ represents real part of $*$.

$\mathcal{I}_{AB}(t)$ and $\mathcal{D}_{AB}(t)$ are evidently the time-ordered correlation functions whereas $\mathcal{F}_{AB}(t)$ is what defines out-of-time-ordered four-point correlation function. 
For the case when $A$ and $B$ are unitary operators, there is a connection between $\mathcal{C}_{AB}(t)$ and $\mathcal{F}_{AB}(t)$
\begin{equation}\label{eq:CandFconnect}
    \mathcal{C}_{AB}^{Unitary}(t)=\left(1 -  \Re\{ \mathcal{F}_{AB}(t) \} \right).
\end{equation}
These four-point correlators show the decay of correlations of the operators that initially commute. Generally, $\mathcal{F}_{AB}(t)$ is a complex number~\cite{Swingle_2024}, but for Hermitian operators  real $\mathcal{F}_{AB}(t)$ is obtained. When the quantum systems are chaotic, the correlations $\mathcal{F}_{AB}(t)$ decay fast as the system loses memory of any perturbation.

In this work, one of our purposes is to study $\mathcal{F}_{AB}(t)$, which can be calculated using an interferometric scheme~\cite{Swingle_2024, Zhang_2019}, for which the name $\mathcal{F}$-OTOC was coined in~\cite{bose2025}.
To measure $\mathcal{F}(t)=\langle A^\dagger{}_t B^\dagger A_t B \rangle$, an interferometric scheme is followed where a control qubit lays out two paths, each facilitating different sequences of operations on a composite initial state of the system and the control qubit. The measurement of the control qubit along the $x$-axis, after the interference, draws out the $\mathcal{F}(t)$.
In \cite{SwingleInfoScram}, this measurement scheme was provided for a unitary evolution. Adding to that,~\cite{Zhang_2019} modified the scheme to incorporate the open system dynamics of GKSL evolution.

An initial composite state is defined where the system state is $\rho_S(0)$ (the form is specified in Sec.~III), and the control qubit state is $\ket{+}_c=\frac{1}{\sqrt{2}}(\ket{0}_c+\ket{1}_c)$.
For a closed system driven by unitary evolutions, the following operations are applied on the initial composite state of the system and the control qubit
\begin{align}
    U_1&= \mathbf{I}_S\otimes \ket{0}\bra{0}_c + B_S \otimes \ket{1}\bra{1}_c, \nonumber \\
    U_2&=e^{-iH_S t}\otimes \mathbf{I}_c, \nonumber \\
    U_3&=A_S\otimes \mathbf{I}_c \nonumber \\
    U_4&=e^{iH_St}\otimes \mathbf{I}_c ,\nonumber \\
    U_5&= B_S \otimes \ket{0}\bra{0}_c + \mathbf{I}_S\ \otimes \ket{1}\bra{1}_c .
\end{align}
After the successive operations of the above unitaries on the initial state, $\rho_f^U$ is obtained, which is then traced over with $\sigma_c^x$ to obtain the $\fotoc$ for closed systems
 \begin{align}\label{unifotoc}
     \mathcal{F}(t,A,B)=&\text{Tr} \left( \sigma_c^x \, \rho_f^U \right) \nonumber \\
     &=\Re~ \text{Tr}(A^{\dagger}(t)B^{\dagger}(0)A(t)B(0)).
 \end{align}

For general open systems, a CPTP  map evolves the local unitary operators in the Heisenberg picture non-unitarily. The forward and backward evolution present in the CPTP is done by the Hamiltonian $H_S$ and $-H_S$. Then the total unitaries for the forward $\mathcal{U}_f$ and backward evolution $\mathcal{U}_b$, respectively, are governed by Hamiltonians $H_f = H_S + H_E + H_{SE}$ and $ H_b = -H_S + H_E + H_{SE} $, respectively, where the bath and the interaction Hamiltonians are kept intact. We denote the forward CPTP map by $\xi_f(t)$ and the backward map by $\xi_b(t)$.
The forward time evolution of operators is done by the adjoint map $\xi_f^{\dagger}(t)$,  such that $A(t)=\xi_f^{\dagger}(t) \cdot A(0)$. Similarly, for the backward evolution driven by $H_b$, we have the backward  CPTP map $\xi_b (t)$ and its adjoint map $\xi_b^{\dagger}(t)$. 
The modified protocol suited to an open system for a general non-Markovian CPTP map $\xi(t)$ (for GKSL evolution, it boils down to $e^{\mathcal{L}t}$, where $\mathcal{L}$ is the Lindbladian superoperator~\cite{Zhang_2019}) is
\begin{align}\label{scheme}
    \mathcal{S}_1&= \mathcal{C}(\mathbf{I}_S\otimes \ket{0}\bra{0}_c + B_S \otimes \ket{1}\bra{1}_c), \nonumber \\
    \mathcal{S}_2&=\xi_f(t)\otimes \mathcal{I}_c, \nonumber \\
    \mathcal{S}_3&=\mathcal{C}(A_S\otimes \mathbf{I}_c) ,\nonumber \\
    \mathcal{S}_4&=\xi_b(t)\otimes \mathcal{I}_c, \nonumber \\
    \mathcal{S}_5&= \mathcal{C}(B_S \otimes \ket{0}\bra{0}_c + \mathbf{I}_S\ \otimes \ket{1}\bra{1}_c ), \nonumber \\
    \rho_f&=\mathcal{S}_5 \cdot \mathcal{S}_4 \cdot \mathcal{S}_3 \cdot \mathcal{S}_2 \cdot \mathcal{S}_1 \cdot \rho_{init},
\end{align}
where $\rho_{init}=\rho_S(0)\otimes\ket{+}_c\bra{+}_c$. 
In $\mathcal{S}_2$, as the CPTP map $\xi_f(t)$ is obtained when the bath is traced out, the identity superoperator $\mathcal{I}_c$ encodes how the control qubit is kept unaffected when the map is acted upon the system and differs from a simple identity matrix operator. Thus, for example, $\mathcal{S}_2$ operation translates to a CPTP map acting on an iteratively updated composite state of the system and the control qubit. The total Hamiltonian that constitutes these CPTP maps has a structure
\begin{align}
    H=H_S\otimes \mathbf{I}_c \otimes \mathbf{I}_E+\mathbf{I}_S \otimes \mathbf{I}_c \otimes H_E + g\sum_{\alpha}(A_S^\alpha \otimes \mathbf{I}_c \otimes A_E^\alpha),
\end{align}
where the empty subspace is added to accommodate the control qubit. Here, $A_S^\alpha$ and $A_E^\alpha$ denote the system and bath operators. The map described in the $\mathcal{S}_2$ can be written in the form
\begin{align}
    \rho_3=\left(\xi_f(t)\otimes \mathcal{I}_c\right)\rho_2=\text{Tr}_B\left[e^{-iHt}\left\{\rho_2\otimes \rho_E(0)\right\}e^{iHt}\right].
\end{align}
Here, $\rho_2$ involves both the system and the control qubit. 
The $\mathcal{I}_c$ accounts for the dimensions of the control qubit being accommodated in the action of the CPTP map.
$\mathcal{C}$ denotes the operation $\mathcal{C}(U) \cdot \rho =U\rho U^{\dagger}$. 
The initial composite state $\rho_{init}$ is subjected to the above sequence of operations, and the OTOC is obtained when the final density matrix $\rho_f$ is traced over with respect to $\sigma^x_c$ ~\cite{Zhang_2019, SwingleInfoScram}, to sort out the off-diagonal terms
 \begin{align}\label{openFotoc}
\mathcal{F}(t, A, B) &:= \text{Tr} \left( \sigma_c^x \, \rho_f \right) \nonumber \\
        &= \Re \, \text{Tr} \left[
            B_S^\dagger  \xi_b(t) \cdot   A_S 
           \left( \xi_f(t) \cdot \left( B_S \, \rho_S(0) \right) \right) 
           A_S^\dagger 
           \right] \nonumber \\
           &= \Re \, \text{Tr} \left[
           \left( \xi_b^\dagger(t) \cdot B_S^\dagger \right) A_S 
           \left( \xi_f(t) \cdot \left( B_S \, \rho_S(0) \right) \right) 
           A_S^\dagger 
           \right].
\end{align}
The details can be found in~\cite{bose2025}. Throughout the paper, we apply the steps outlined in Eq.~\eqref{scheme} and perform the trace operation as given in Eq.~\eqref{openFotoc} numerically to calculate the interferometric OTOC.

\subsubsection{Corrected $\mathcal{F}$-OTOC}
For open systems, the $\mathcal{F}$-OTOC shows information scrambling in the system along with the open system effects.. The scrambling is the consequence of the non-commutativity between $A_S(t)$ and $B_S$ operators, and the  $\left( \xi_b^\dagger(t) \cdot B_S \right)$ term brings about the open system effects. The effect of pure scrambling is brought out when $\mathcal{F}$-OTOC is corrected by factoring out the effect of dissipation. In \cite{SwingleInfoScram}, a corrected OTOC was proposed by dividing $\mathcal{F}(t, A, B)$  by $\mathcal{F}(t, I, B)$ because of the fact that $\mathcal{F}(t, I, B)$ contains purely dissipative effects as $I$ and $B$ always commute. The definition of the corrected $\mathcal{F}$-OTOC is 
\begin{align}\label{CorrectedOTOC}
    \mathcal{F}_c(t)=\frac{\mathcal{F}(t,A,B)}{\mathcal{F}(t,I,B)}.
\end{align}

\subsection{Bipartite OTOC}
One of the vital facets of scrambling of quantum information is the non-commutativity of local operators. To proceed with the bipartite OTOC, we constrain ourselves to unitary local operators and assume the state of the system to be a maximally mixed state $\rho = \frac{\mathbb{I}}d$.
The exponential profile of $C_{AB}(t)$, Eq. (\ref{commusquare}), hints at the chaotic nature of a quantum system. This has been verified in systems with semiclassical limits or systems with multiple degrees of freedom~\cite{Hirscheatomfield}.
The bipartite OTOC examines the evolution of the non-commutativity of two unitary operators based on two partitions of a Hilbert space. 
The choice of those unitary operators is essential for OTOCs. This reliance on the specific choice of unitary operators is intended to be eliminated by the concept of bipartite OTOC. A more generic variant of the OTOC is produced by averaging across all potential random unitary operators because not all of the unitary operators are equally sensitive to chaotic behavior~\cite{22nofromzanardipaper,30nofromzanardipaper,31nofromzanardi,32nozanardipaper,33nofromzanardipaper,34fromzanardipaper,35nofromzanardipaper}.

A finite-dimensional composite Hilbert space is considered with two bipartitions $\mathcal{A}$ and $\mathcal{B}$ of dimensions $d_A$ and $d_B$, respectively. $\mathcal{H}=\mathcal{H}_A \otimes \mathcal{H}_B$ is the total Hilbert space of the system, having dimension $d=d_Ad_B$. The commutator square $C_{AB}(t)$ in Eq. \eqref{commusquare} and consequently Eq. \eqref{eq:CandFconnect}, is Haar averaged over all independent unitary operators $A$ and $B$, which belong to  $\mathcal{A}$ and $\mathcal{B}$ partitions of the system, respectively. This yields,
\begin{align}\label{eq:1st average}
    G(t)&= 1-\frac{1}{d}{\Re}\int{d\mu(A)d\mu(B){\rm Tr}\left[A^{\dagger}(t) B^{\dagger} A(t) B\right]} \nonumber \\
    &=1-\int{d\mu(A)d\mu(B)\mathcal{F}(A,B,t)}.
\end{align}
The $d\mu(A, B)$'s are the differential elements of the unitary matrices $A$ and $B$.
Here,
\begin{align}\label{eq_Fbip}
    \mathcal{F}(t)=\text{Tr}\left(A^{\dagger}(t)B^{\dagger}A(t)B\frac{I}{d}\right),
\end{align}
and the trace is done with respect to the maximally mixed states.
The averaging over random unitary matrices is done by using some Haar averaging identities that bring about swap operators based on an extension of the system Hilbert space $\mathcal{H}$ by its replica $\mathcal{H'}$~\cite{Zanardiopenbipotoc1, Bose_2024}. The bipartite OTOC for the unitary evolution of the system is,
\begin{equation}\label{eq:5}
    G(t)=1-\frac{1}{d^2}{\rm Tr}(S_{AA^{\prime}}U_t^{\otimes2}S_{AA^{\prime}}U_t^{\dagger \otimes2}),
\end{equation}
Where $U_t=e^{-iHt}$ is the unitary operator and $H$ is the Hamiltonian of the composite system $(\mathcal{A}\mathcal{B})$, respectively.
The swap operators have the following properties,
\begin{equation}
       S=S_{AA^{\prime}}S_{BB^{\prime}}=S_{BB^{\prime}}S_{AA^{\prime}}
\end{equation}
and  $S^2=S_{AA^{\prime}}^2=S_{BB^{\prime}}^2=\mathbb{I}$. They operate on $\mathcal{H}\otimes\mathcal{H}'=\mathcal{H}_A \otimes \mathcal{H}_B \otimes \mathcal{H}_A' \otimes \mathcal{H}_B'$.
For an open system, where the system evolves not under a unitary map but rather a CPTP map $\mathcal{E}$ and the operators evolve under its adjoint map $\mathcal{E}^{\dagger}$, the bipartite OTOC is~\cite{Zanardiopenbipotoc1},
    \begin{align}
        G(\mathcal{E}^{\dagger})&\coloneqq \frac{1}{2d} \mathbb{E}_{A,B}[C_{A,B}(t)] =\frac{1}{2d} \mathbb{E}_{A,B}||[\mathcal{E}^{\dagger}(A),B]||^2_2 \nonumber \\
        &=\frac{1}{2d} \mathbb{E}_{A,B} {\rm Tr} \{ [\mathcal{E}^{\dagger}(A),B]^{\dagger}[\mathcal{E}^{\dagger}(A),B]\} . 
        \label{channeltr}
    \end{align}
The symbol $\mathbb{E}_{A, B}$ denotes Haar averaging with respect to all possible unitary operators in the Hilbert space.
Using the swap operators $S_{AA'}$ and $S$, the bipartite OTOC for open quantum systems is obtained as~\cite{Zanardiopenbipotoc1, Bose_2024} 
\begin{align}\label{Openbotoc}
    G(\mathcal{E}^{\dagger})=&\frac{1}{d} \bigg\{ {\rm Tr}\big( S\mathcal{E}^{\dagger \otimes 2} \frac{S_{AA^{\prime}}}{d_A} \big)- \frac{1}{d}{\rm Tr} \big( S_{AA^{\prime}} \mathcal{E}^{\dagger \otimes 2} S_{AA^{\prime}} \big) \bigg\}, \nonumber \\
   =&\frac{1}{d^2}{\rm Tr} \{ (Sd_B-S_{AA^{\prime}}) \mathcal{E}^{\dagger \otimes 2} S_{AA^{\prime}} \}.
\end{align}
To calculate the bipartite OTOC, only the adjoint map $\mathcal{E}^{\dagger}$ and the swap operators are needed. Here, the models are defined by their Hamiltonians, which are used to evolve the system. This evolution is characterised by the map from which
the superoperator involved in the calculation of Eq.~\eqref{Openbotoc} can be extracted.
The CPTP adjoint map $\mathcal{E}^{\dagger}$ and the swap operators $S,~S_{AA'}$ are obtained numerically using the computational basis vectors following the prescriptions elaborated in Appendix A and B, respectively.

\subsection{The models}
\subsubsection{A spin-chain interacting with LMG bath}
Here, we discuss a model of spin-chain interacting with a finite Lipkin-Meshkov-Glick (LMG) bath~\cite{Vidal_LMG}. The spins interact uniformly with all the spins of the LMG bath, and they interact with each other. The Hamiltonian of the total system is given by
\begin{align}
    H = H_S + H_E^{LMG} + H_{SE}^{LMG},
\end{align}
where $H_S$ is the system Hamiltonian, constituting central spins
\begin{align}
    H_S = \sum_{j = 1}^{N_S}\omega_j\sigma_j^z + \sum_{j = 1}^{N_S-1} J_{j}\sigma^z_{j}\sigma^z_{j+1},
\end{align}
where $N_S$ is the number of spins in the system, $\omega_j$'s are their transition frequencies, and $J_{j}$ is the interaction strength between qubits $j$ and $j+1$. The system Hamiltonian depicts the Ising model type of nearest neighbour interaction.  
The environment is the isotropic LMG bath~\cite{Vidal_LMG}, where all qubits interact with each other. The environment Hamiltonian is given by~\cite{Hou2016}
\begin{align}
    H_E^{LMG} = \frac{\lambda}{N}\sum_{i<j}^N\left(\sigma^x_i\sigma^x_j + \sigma^y_i\sigma^y_j\right) + \omega_c\sum_{i=1}^N \sigma^z_i,
\end{align}
where $N$ is the number of qubits in the environment, $\lambda$ is the interaction strength between the bath qubits, and $\omega_c$ is their transition frequency. By using the collective angular momentum operators $J^k_N = \frac{1}{2}\sum_{j=1}^N\sigma^k_j$, the bath Hamiltonian can be rewritten as
\begin{align}
    H_E^{LMG} = \frac{2\lambda}{N}\left(J_N^x\otimes J_{N-1}^x + J_N^y\otimes J_{N-1}^y\right) + 2\omega_c J^z_N.
\end{align}
Further, the interaction Hamiltonian $H_{SE}^{LMG}$ is given by
\begin{align}
    H_{SE}^{LMG} = \frac{\tilde \lambda}{\sqrt{N}}\sum_{j =1}^{N_S}\left(\sigma^x_jJ^x_N+\sigma^y_jJ^y_N\right),
\end{align}
where $\tilde\lambda$ is the interaction strength between the system qubits and the bath qubits. Throughout the paper, we take $\tilde\lambda = \lambda$. It can be noted that for $N_S=1$, the model reduces to a central spin model~\cite{devvrat_central_spin_dynamics} with interacting bath spins.

\subsubsection{Tilted field Ising model interacting with an anisotropic spin chain bath}
We also study a model where a tilted field Ising model (TFIM)~\cite{TiltedMfieldofarul, Bose_2024} interacts with an anisotropic spin chain bath~\cite{Wei2016}. The Hamiltonian for this model is given by
\begin{align}
    \tilde H = \tilde H_S + \tilde H_E + \tilde H_{SE},
\end{align}
where the system Hamiltonian $\tilde H_S$ is given by
\begin{align}\label{TFIM_hamil}
    \tilde H_S = \mathcal{B}\sum_{i = 1}^{\tilde N_S}\left\{\sin(\theta)\sigma^x_i + \cos(\theta)\sigma^z_i\right\} +\mathcal{J}\sum_{i=1}^{\tilde N_S-1}\sigma^z_i\sigma^z_{i+1},
\end{align}
where $\mathcal{B}$ is the strength of the magnetic field, $\mathcal{J}$ is the coupling strength between the system spins, and $\theta$ is the angle of the tilt between the magnetic field and the spin axis. The environment interacts with the last qubit ($\tilde N_S^{th}$-qubit) via the Hamiltonian
\begin{align}\label{ani_interac}
    \tilde H_{SE} = g\left(\sigma^x_{\tilde N_S}\sum_{l = 1}^M\sigma^x_l + \sigma^y_{\tilde N_S}\sum_{l = 1}^M\sigma^y_l\right),
\end{align}
where the coupling strength is $g$ and $M$ is the number of spins in the environment. The environment is composed of a spin chain with an anisotropic interaction, whose Hamiltonian is given by
\begin{align}
    H_E = \sum_{l=1}^M\left[\left\{\frac{1+\gamma}{2}\right\}\sigma^x_l\sigma^x_{l+1} + \left\{\frac{1-\gamma}{2}\right\}\sigma^y_l\sigma^y_{l+1} + \lambda_z\sigma^z_l\right].
\end{align}
Here, the parameter $\lambda_z$ denotes the strength of the magnetic field, and $\gamma$ is the anisotropic parameter. The value of $\gamma=1$ and $\gamma=0$ corresponds to the Ising and Heisenberg XX models, respectively. Also, for $l=M$-th spin in the bath, the $l+1$-th spin is the $l=1$-th spin, making it a closed chain. 

\section{$\mathcal{F}-$OTOC and bipartite OTOC for the spin-chain model interacting with the LMG bath}\label{sec_III}
\begin{figure}
    \centering
    \includegraphics[width=1\columnwidth]{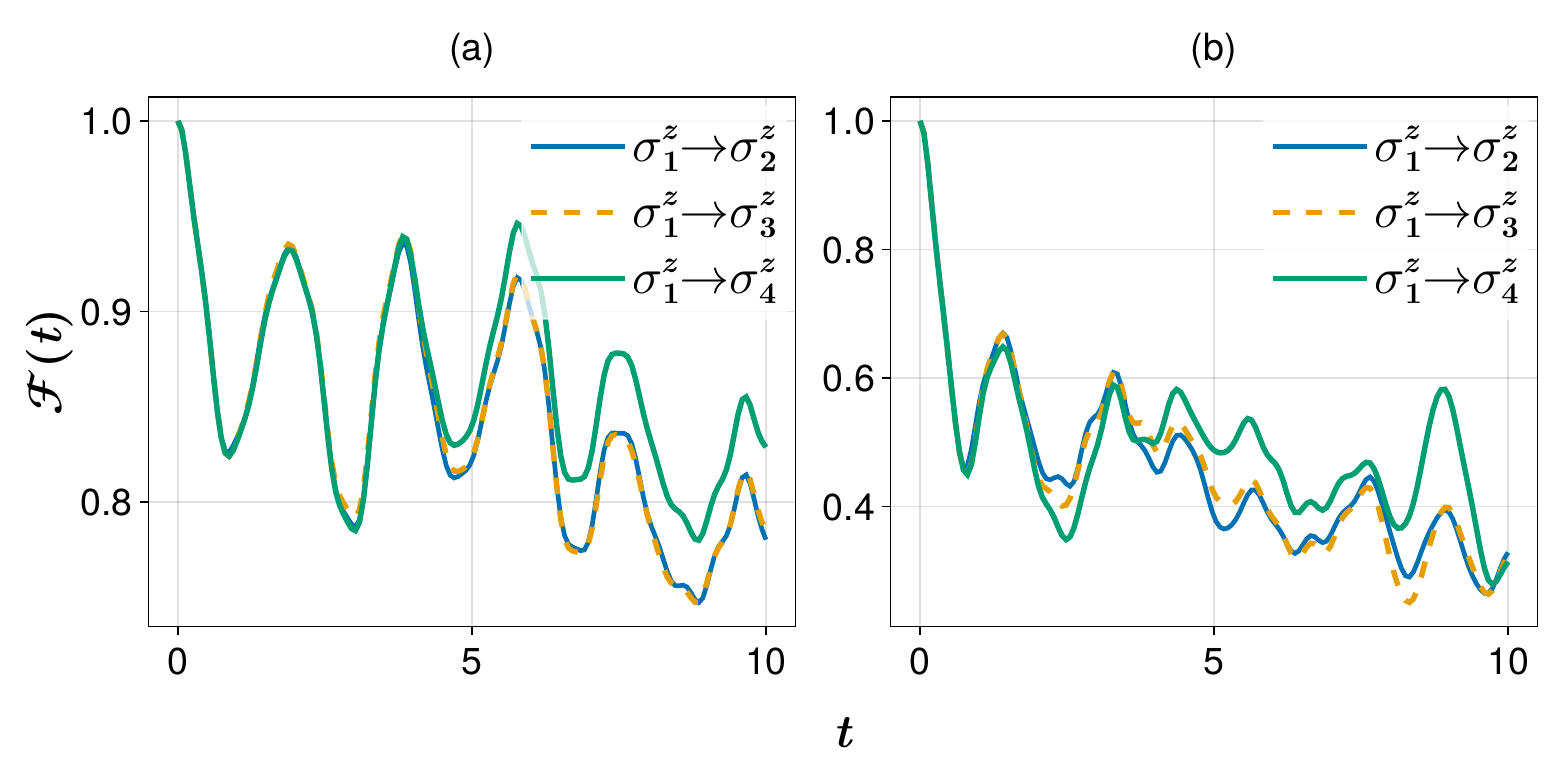}
    \caption{The $\mathcal{F}(t)$ is plotted for the system of four spins under the LMG bath. Information scrambling is observed from the first spin, denoted by $\sigma_z^1$, to the other spins as depicted by the direction of the arrow. In (a), the $\mathcal{F}(t)$ is when $\lambda=0.5$ and (b) is when $\lambda=1$, where $\lambda$ is the constant of coupling to the bath. The other parameters are, $J_j=0.5$, $\omega_0=2$, $\omega_c=4, N = 5$, and $T=10$.}
    \label{fig:FOTOC_four_spins}
\end{figure}

\begin{figure}
    \centering
    \includegraphics[width=1\columnwidth]{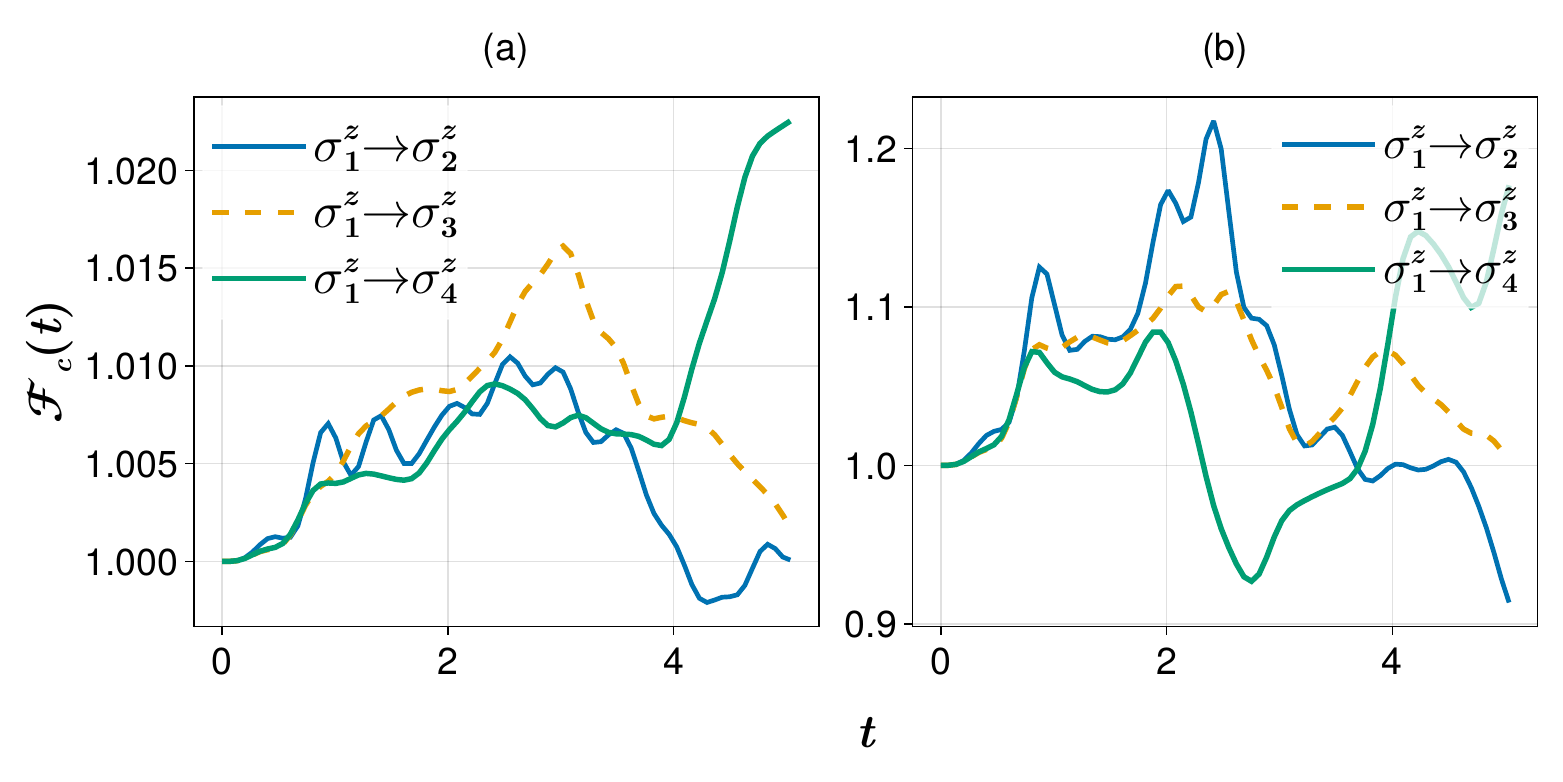}
    \caption{The plot for $\mathcal{F}_c(t)$ when the system is of four spins under the LMG bath. Information scrambling is observed from the first spin, denoted by $\sigma_z^1$, to the other spins as depicted by the direction of the arrow. In (a), the $\mathcal{F}_c(t)$ is when $\lambda=0.5$, and (b) is when $\lambda=1$, where $\lambda$ is the constant of coupling to the bath. The other parameters are, $J_j=0.5$, $\omega_0=2$, $\omega_c=4, N = 5$, and $T=10$.}
    \label{fig:Corr_FOTOC_four_spins}
\end{figure}

To analyze information scrambling, operator growth from the first spin to the other spins is observed. The initial states of the system and the bath are taken to be $\rho_S(0) = (\ket{\psi(0)}_S\bra{\psi(0)}_S)^{\otimes N_S}$, where $\ket{\psi_S(0)} = \frac{\sqrt{3}}{2}\ket{0} + \frac{1}{2}\ket{1}$, and $\rho_E(0) = e^{-H^{LMG}_E/T}/ {\rm Tr}\left[e^{-H^{LMG}_E/T}\right]$, respectively, where $T$ is the temperature of the bath. The local operators $A_S$ and $B_S$ are taken as Pauli $\sigma_z$ throughout the paper.
In Fig.~\ref{fig:FOTOC_four_spins}, the $\mathcal{F}-$OTOC [$\mathcal{F}(t)$, Eq.~\eqref{openFotoc}] is shown for a system of four spins under the LMG bath, i.e., when $N_S=4$. 
The comparison between Fig.~\ref{fig:FOTOC_four_spins}(a) and (b) shows that a higher coupling to the environment makes way for fast scrambling, as evident from the decay of $\fotoc$ at small times to much lower values when $\lambda$ is higher. 
Since each spin is interacting with the LMG bath, the information light cone \cite{SwingleInfoScram, Zhang_2019} is not preserved. The open system effects take precedence over the pure scrambling of information in the system. The corrected $\mathcal{F}-$OTOC, Eq.~\eqref{CorrectedOTOC}, as a measure of pure information scrambling constructed in a way to partially get rid of the open system effects, is depicted in Fig.~\ref{fig:Corr_FOTOC_four_spins}. However, even the $\mathcal{F}_c(t)$'s don't show any clear light cone phenomena. In~\cite{bose2025}, it was observed that the corrected $\mathcal{F}-$OTOC only partially retrieves the light-cone in the case of a radiation bath globally connected to a four-spin Ising system. The fact that the corrected OTOC does not retrieve the light-cone to any degree here could be ascribed to the greater number of channels available to the spin-chain for the dissipation of information by virtue of the inter-qubit interactions among the spins of the LMG bath. 

\subsection{Comparison of $\mathcal{F}-$OTOC and bipartite OTOC in case of an Ising chain of two spins in an LMG bath}
\begin{figure}
    \centering
    \includegraphics[width=1\columnwidth]{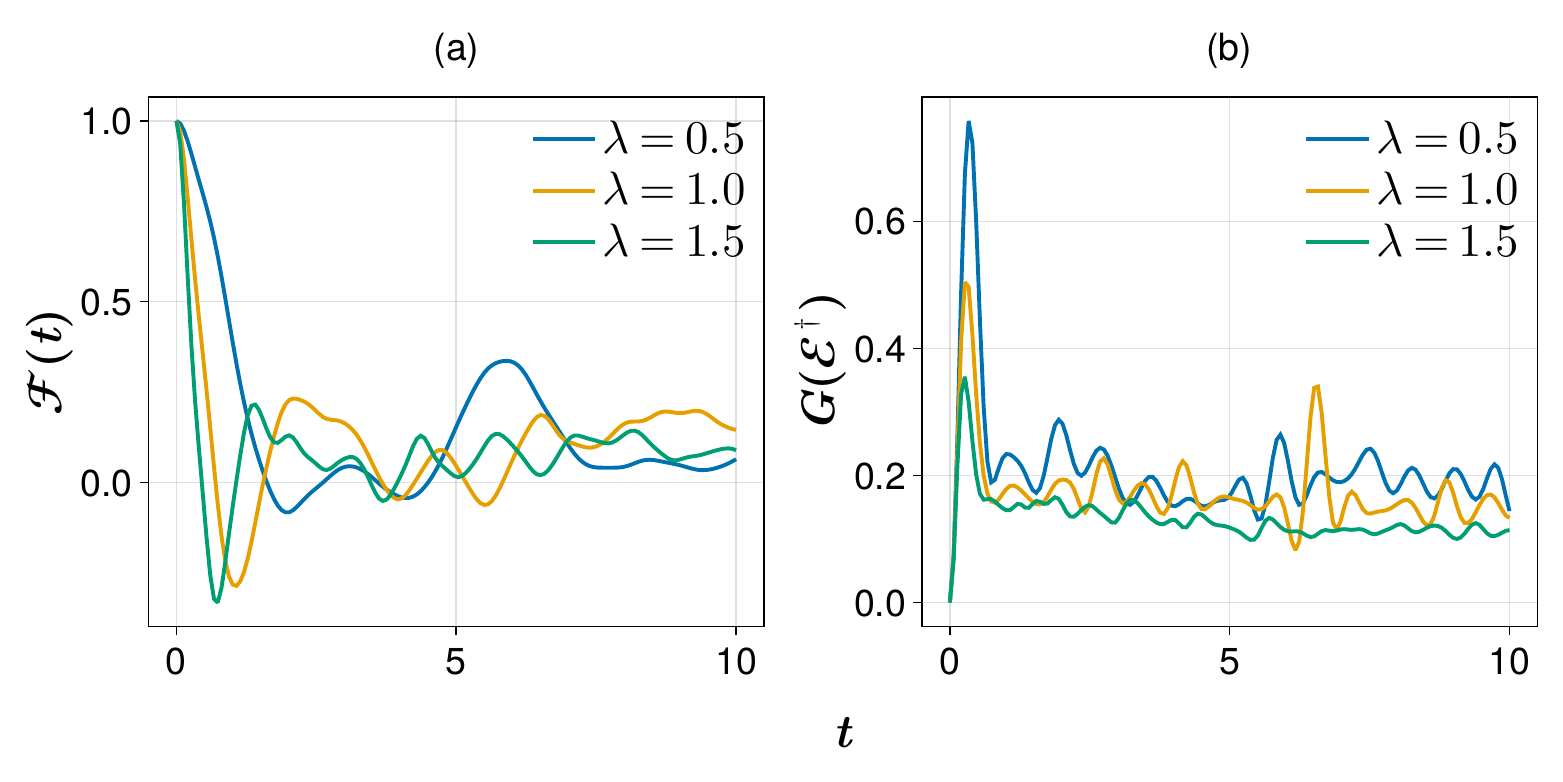}
    \caption{$\mathcal{F}(t)$ in (a) and $G(\mathcal{E}^{\dagger})$ in (b) are plotted for two spins surrounded by an LMG bath. The bipartitions needed for calculating the bipartite OTOC, $G(\mathcal{E}^{\dagger})$, are taken as two two-qubit Hilbert spaces. The variations are shown for different coupling constants with the bath $\lambda$. The other parameters are, $J_j=0.5$, $\omega_0=2$, $\omega_c=4, N = 10$, and $T=10$.}
    \label{fig:Fotoc_Botoc_varyLambda}
\end{figure}

\begin{figure}
    \centering
    \includegraphics[width=1\columnwidth]{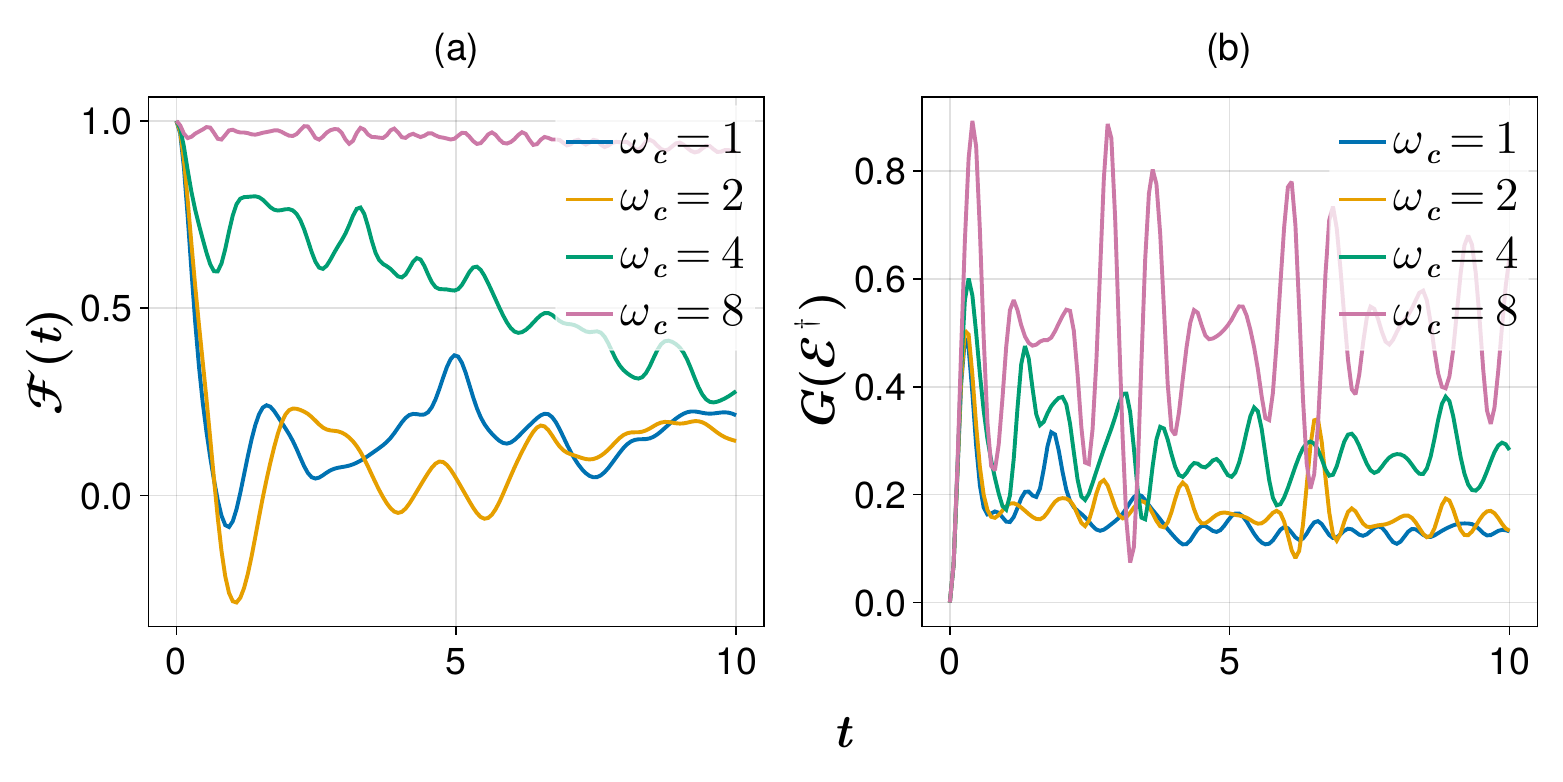}
    \caption{$\mathcal{F}(t)$ in (a) and $G(\mathcal{E}^{\dagger})$ in (b) are plotted for two spins surrounded by an LMG bath. The variations are shown for different bath frequencies $\omega_c$. The other parameters are, $J_j=0.5$, $\omega_0=2$, $\lambda=1$, $N = 10$, and $T=10$.}
    \label{fig:Fotoc_Botoc_varywc}
\end{figure}

\begin{figure}
    \centering
    \includegraphics[width=1\columnwidth]{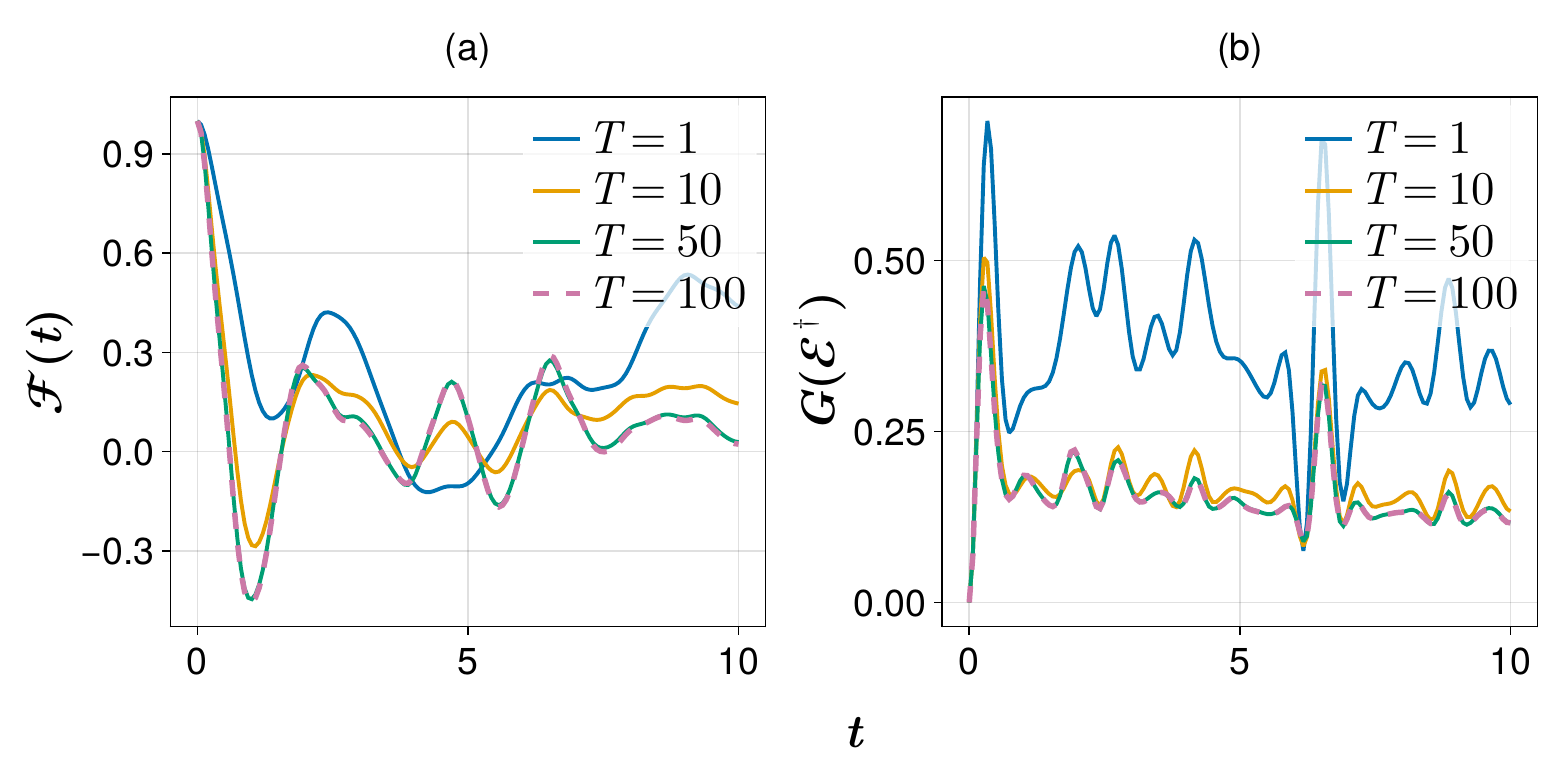}
    \caption{$\mathcal{F}(t)$ in (a) and $G(\mathcal{E}^{\dagger})$ in (b) are plotted for two spins surrounded by an LMG bath. The variations are shown for different temperatures of the bath $T$. The other parameters are $J_j=0.5$, $\omega_0=2$, $\lambda=1$, $N = 10$, and $\omega_c=2$.}
    \label{fig:Fotoc_Botoc_varyT}
\end{figure}

To compare the $\mathcal{F}-$OTOC  and bipartite OTOC, we work on a system of two spins ($N_S=2$) so that the bipartition of the system needed to obtain $G(\mathcal{E}^{\dagger})$, Eq.~\eqref{Openbotoc}, is according to the dimensions of the two local operators $A_S$ and $B_S$. Since each of the two local operators of two spins has dimension two, we divide the system into two two-dimensional partitions. It is to be noted that the four-point correlation function $\mathcal{F}(t)$ starts from the state of being fully correlated, i.e., $\mathcal{F}(t)=1$, and then decreases as the system evolves under the CPTP map, losing correlations by the combined effects of information scrambling and dissipation. In contrast, the bipartite OTOC starts from zero because the initial local operators commute. This can also be seen by examining Eqs.~\eqref{eq:1st average} and~\eqref{eq_Fbip} for $t=0$. 
As we have seen from the definition of bipartite OTOC, the Haar averaging of the commutator of two local operators is done with respect to the maximally mixed state. Here, we also employ the same initial state to calculate the $\fotoc$ in order to elicit a proper comparison between these two quantities.
In Fig.~\ref{fig:Fotoc_Botoc_varyLambda}, we see the plots for $\mathcal{F}(t)$ and $G(\mathcal{E}^{\dagger})$ for different values of the coupling constant $\lambda$  of the LMG bath with the two-spin system. As the coupling gets stronger, the information of a given measurement-induced perturbation scrambles faster. This phenomenon is manifested in the slope of the short-time decay of $\mathcal{F}(t)$ and the height of the short-time peak of the $G(\mathcal{E}^{\dagger})$. For a higher coupling constant, the $\fotoc$'s decay is steeper and the $G(\mathcal{E}^{\dagger})$'s peaks are shorter.
In Fig.~\ref{fig:Fotoc_Botoc_varywc}, it is shown that in case of a sluggish bath~\cite{Breuer2007}, i.e., when the bath frequency is much higher than that of the system; $\omega_c\gg\omega_0$, the scrambling is significantly low as the $\mathcal{F}(t)$ barely decays and the $G(\mathcal{E}^{\dagger})$ is nearly harmonic.
In Fig.~\ref{fig:Fotoc_Botoc_varyT}, it is seen that a higher temperature leads to faster scrambling as hinted by the steeper decay of $\fotoc$ but lower height of the short-time peak of $G(\mathcal{E}^{\dagger})$. Interestingly, variation of the coupling constant $\lambda$ has similar effects on $\fotoc$ and on $G(\mathcal{E}^{\dagger})$ as the variation of $T$. At higher temperatures, both of the quantities saturate to specific profiles. The $\fotoc$ is more sensitive to the intervals of high temperatures, whereas $G(\mathcal{E}^{\dagger})$ quickly reaches saturation.

A transition from non-Markovian to Markovian dynamics, for the open quantum system models studied here, is not clear. For the spin-chain model interacting with a bosonic bath, an analysis of bipartite OTOC has been done in~\cite{Bose_2024} recently, making use of the Gorini-Kossakowski-Sudarshan-Lindblad master equation (Markovian in nature). In the non-Markovian case, the decay of the bipartite OTOC is faster, and the fluctuations in the tail are haphazard, while in the Markovian case, the oscillatory decay of the bipartite OTOC has been observed to be smoother.

\section{$\mathcal{F}-$OTOC for TFIM interacting with an anisotropic spin chain}\label{Sec_IV}
Here, we discuss the $\mathcal{F}-$OTOC for the TFIM interacting with an anisotropic spin chain. 
\begin{figure}
    \centering
    \includegraphics[width=1\linewidth]{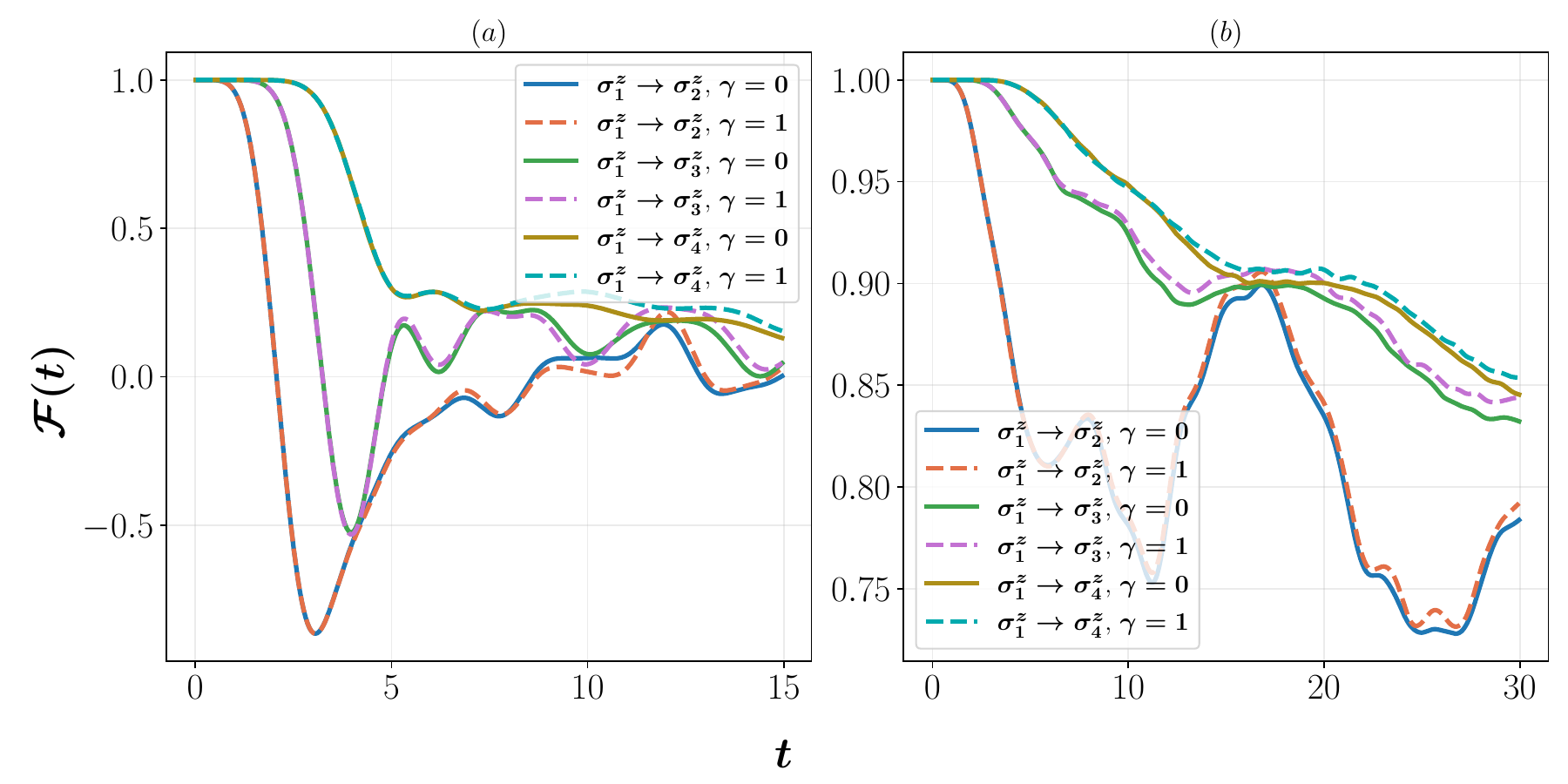}
    \caption{Variation of $\mathcal{F}-$OTOC $\mathcal{F}(t)$ for the TFIM interacting with the anisotropic spin chain. (a): $\theta = \pi/2$, and (b): $\theta = \pi/8$. The parameters are: $\mathcal{B} = \mathcal{J} = 0.5, T = 10, M=6, \lambda_z = 1.0$, and $g = 0.5$. The initial states of the system and the bath are taken to be $\ket{\psi(0)}_S^{\otimes 4}$, where $\ket{\psi}_S = \frac{\sqrt{3}}{2}\ket{0} + \frac{1}{2}\ket{1}$ and $\rho_B(0) = e^{-\tilde H_E/T}/{\rm Tr}\left[e^{-\tilde H_E/T}\right]$.}
    \label{fig_fotoc_phase_transition_model}
\end{figure}
In Fig.~\ref{fig_fotoc_phase_transition_model}, plots for both the non-trivially integrable model at $\theta=\pi/2$ and the non-integrable model at $\theta=\pi/8$ are shown, where $\theta$ is the angle of tilt of the magnetic field, Eq.~\eqref{TFIM_hamil}. The nature of interaction in Eq.~\eqref{ani_interac} is such that the bath is connected to only the last spin, i.e., the $\tilde{N}_S^{th}$ spin. This model is different from the previous model of a spin-chain in an LMG bath, where each spin interacts equivalently with the bath. Due to this distinction, the TFIM shows a key feature of information scrambling - the information light cone. The OTOCs observing scrambling between two far-away spins take more time to start decaying compared to those of the nearer spins. The extent of the initial $\mathcal{F}(t)=1$ is the marker for greater distances between the given two spins. It is observed that the length of $\mathcal{F}(t)=1$ linearly increases with the lattice distance between two spins. This elaborates the fact that the information caused by a perturbation propagates with a finite speed and travels through the lattice by virtue of the nearest-neighbor interaction and is the remnant of the light cone phenomenon, which becomes apparent in a thermal map of $\mathcal{F-}$OTOCs~\cite{Zhang_2019}.
In Figs.~\ref{fig_fotoc_phase_transition_model}(a) and (b), the initial values of $\mathcal{F}(t)$ remain unity for some time. This time is longer when the distance between the two local operators is greater. This brings out the phenomenon of the information light cone manifesting in Ising spin chains. It is interesting to see the preservation of the information light cone in both the non-trivially integrable model ($\theta=\pi/2$) and the non-integrable model ($\theta=\pi/8$) shown in Figs.~\ref{fig_fotoc_phase_transition_model}(a) and (b), respectively.

\section{$\mathcal{F}-$OTOC for the LMG model}\label{Sec_V}
Up to this point, the LMG model was presented in the guise of a bath. Since it is an interesting model and has several phases, it would be of interest to investigate it from the perspective of information scrambling.
This motivates us to calculate $\mathcal{F}(t)$ for the LMG model in a closed system scenario, involving unitary evolution for calculating $\mathcal{F}-$OTOC, Eq.~\eqref{unifotoc}. The LMG model has global interaction among the spins. 
\begin{equation}
    H^{LMG} = \frac{\lambda}{N_S}\sum_{i<j}^{N_S}\left(\sigma^x_i\sigma^x_j + \gamma\sigma^y_i\sigma^y_j\right) + \omega_c\sum_{i=1}^{N_S} \sigma^z_i,
\end{equation}
is the Hamiltonian for the LMG model, and $\gamma$ is the anisotropy parameter, while $\omega_c$ is the frequency of individual spins. The model has been well studied in the literature~\cite{Vidal_LMG, LIPKIN1965188, MESHKOV1965199, GLICK1965211, Vidal_LMG}. There are multiple phases possible for this model. The first one is the symmetric phase ($\omega_c\ge 1$), where the ground state is unique. The other is the broken phase ($0\le\omega_c<1$ and $\gamma\ne 1$), where the ground state is two-fold degenerate. The model is in the isotropic phase when $\gamma=1$, where the ground state is infinitely degenerate.
\begin{figure}
    \centering
    \includegraphics[width=1\linewidth]{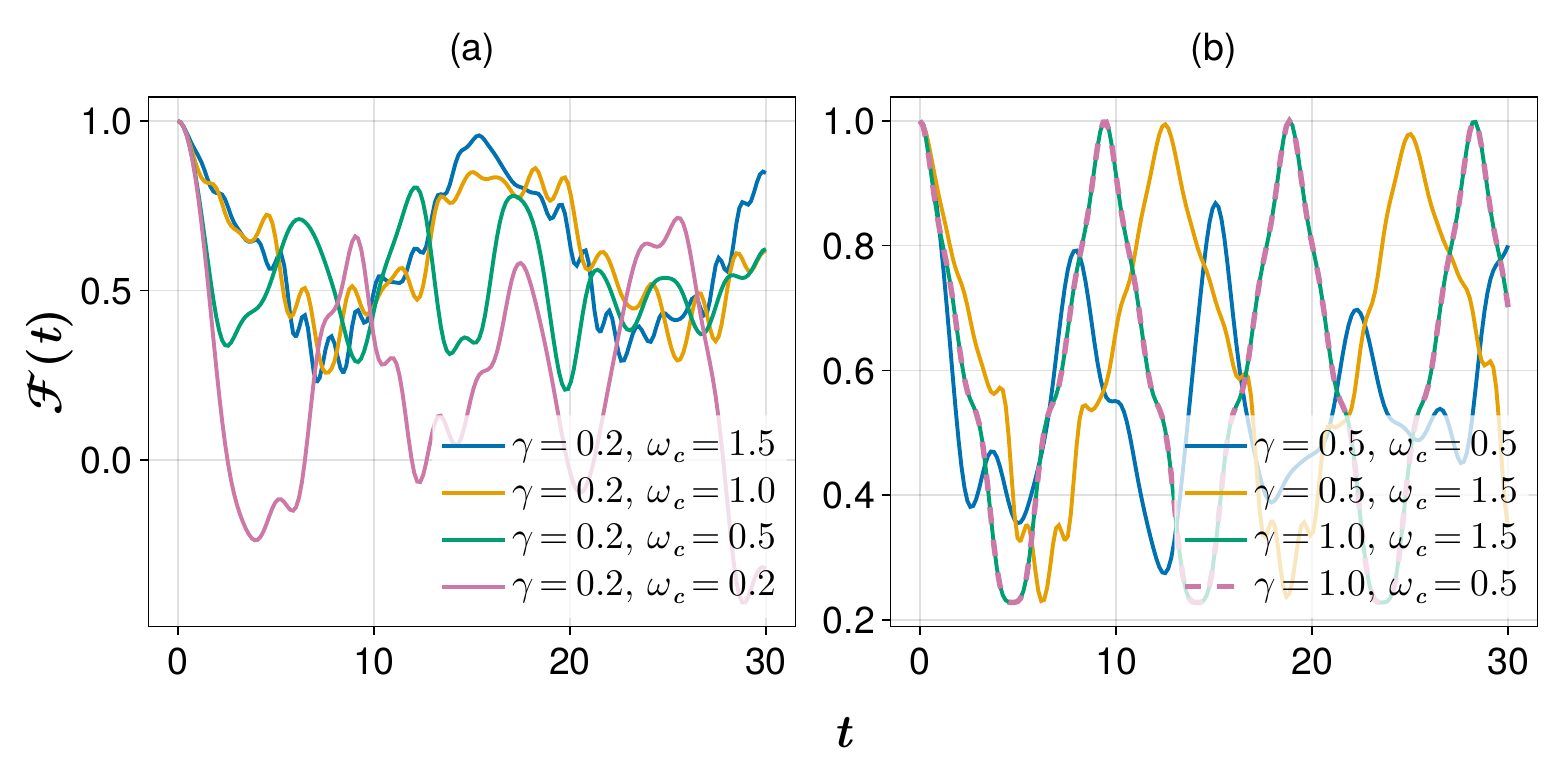}
    \caption{$\mathcal{F}-$OTOC for the LMG model for variation of parameters $\gamma$ and $\omega_c$ when $A_S$ and $B_S$ are both $\sigma_z$ taken at different spin locations. Other parameters are $N_S=6$, $\lambda=1$}
    \label{fig:Fotoc_LMG_SzSz}
\end{figure}
\begin{figure}
    \centering
    \includegraphics[width=1\linewidth]{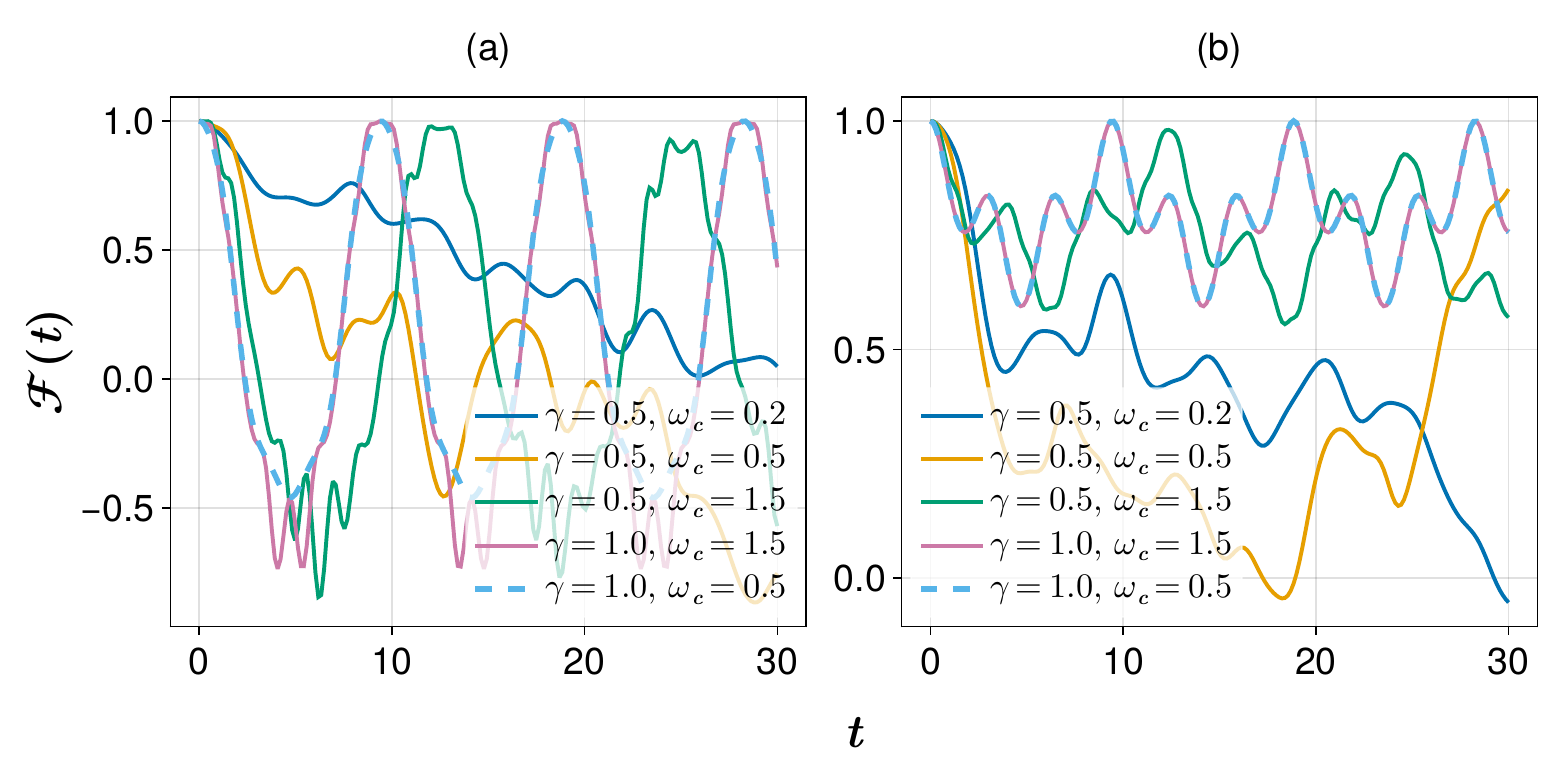}
    \caption{$\mathcal{F}-$OTOC for the LMG model for variation of parameters $\gamma$ and $\omega_c$. In (a), $A_S$ and $B_S$ are both $\sigma_x$'s, and in (b), $A_S=\sigma_z$ while $B_S=\sigma_x$ taken at different spin locations. Other parameters are $N_S=6$, $\lambda=1$.}
    \label{fig:Fotoc_LMG_SxSx_SxSz}
\end{figure}

The $\mathcal{F}(t)$'s in Figs.~\ref{fig:Fotoc_LMG_SzSz} and~\ref{fig:Fotoc_LMG_SxSx_SxSz} show pure information scrambling. 
We observe that for $\gamma=1$, the $\fotoc$ is periodic irrespective of the value of $\omega_c$, indicating integrability in the system and the absence of chaos. When $0\leq\omega_c<1$ and $\gamma\ne 1$, i.e., when the system is in the broken phase, the $\fotoc$ decays irregularly and shows non-periodic behavior, hinting at chaos in the system. This confirms that the LMG model undergoes a phase transition in this range. The LMG model is constructed in such a way that each spin interacts with every other spin (Global interaction). This is the reason why the $\fotoc$ doesn't depend on the distance between the sites where the local operations $A_S$ and $B_S$ are applied. 

In the isotropic phase of the system, i.e., $\gamma=1$, no scrambling of information is observed since we have periodic OTOCs irrespective of $\omega_c$ values. In the $\gamma<1$ and $\omega_c\geq1$ region (symmetric phase), very little or no scrambling is observed. A weaker $\gamma$ allows for a little scrambling. When in the broken-symmetry phase ($0\leq \omega_c <1$), the OTOCs decay significantly, hinting that the information scrambles and this phase of the system is chaotic in nature. The interplay between the chaotic nature of the dynamics of the LMG Hamiltonian with its different phases is noteworthy. It can be inferred from the above analysis that only in the broken symmetry phase, the LMG model is a fast scrambler of information and is chaotic.

\section{Conclusions}\label{sec_conclusions}
In this work, facets of information scrambling in spin-chain models were explored. To this effect, the following models were considered. The first one was the Ising spin chain interacting with a Lipkin-Meshkov-Glick (LMG) type bath, while the second was a tilted field Ising model (TFIM) interacting with an anisotropic spin chain bath at one extremity. The models are chosen to observe the different effects they have on the saturation time and the saturation value of information scrambling measures. To characterize the information scrambling, interferometric OTOC ($\mathcal{F}-$OTOC) and bipartite OTOC were studied and compared.

In the Ising chain model interacting with the LMG bath, decay of $\mathcal{F}-$OTOC was observed, which could be attributed to the cumulative effects of fast scrambling of information and non-Markovian open system effects. No light cone phenomena were observed. This could be expected from the nature of interaction with the bath, as it impacts each individual spin of the system. In the same model, for a system of two spins under the effect of the LMG bath, $\mathcal{F}-$OTOC and bipartite OTOC were calculated taking the two local spin operators as the bipartite system, and their results were compared. We found that these two quantities have their own way of indicating key features of the dynamics. For the TFIM interacting with the anisotropic spin-chain bath, $\mathcal{F}-$OTOC showed a clear light cone as it is connected by the last spin to the bath, confirming the existence of ballistic scrambling of information through the spins. Motivated by the interesting features present in the LMG model itself, such as the categorization between symmetric and broken phases, we explored $\mathcal{F-}$OTOC for the closed LMG model. The LMG model is characteristically a system with global interactions among the spins, and when studied from the perspective of information scrambling, the $\mathcal{F}-$OTOC indicated chaos only in the broken phase of the model and showed periodic behavior otherwise. In summary, the saturation value and saturation time manifested their own signatures in the bipartite and the $\mathcal{F}-$OTOC for different models. The slopes of the decay in $\mathcal{F}-$OTOC and the height of the initial rise in the bipartite OTOC characterized the speed of scrambling, and hence the chaotic dynamics, in different parameter regimes.
This study provides a deeper insight into the chaotic behavior of the spin-chain and the LMG model using interferometric OTOC and bipartite OTOC as measures of information scrambling, and explores a connection between information scrambling and phase transition. 

\appendix
\section{Calculation of $\mathcal{E}^{\dagger}$}
The action of an adjoint CPTP  (completely positive trace-preserving) map on an initial system operator $A_S(0)$ is the following
\begin{align}
    A_S(t)&=\Edag[A_S(0)] \nonumber \\
    &=\Edag\sum_{ij}A_{ij}\ket{i}\bra{j} =\sum_{ij}A_{ij}\Edag(\ket{i}\bra{j}).
\end{align}
On vectorising this equation, we obtain
\begin{align}\label{rhoStEdag}
    \ket{A_S(t)}\rangle=\sum_{ij}A_{ij}\ket{\Edag(\ket{i}\bra{j})}\rangle.
\end{align}
Here, we find a matrix $\mathcal{S}$ that represents the above map $\Edag$ in a superoperator space. To this end, we can rewrite the above equation as
\begin{align}
    \ket{A_S(t)}\rangle=\mathcal{S}\ket{A_S(0)}\rangle.
\end{align}
On decomposing the elements in the above equation, we find
\begin{align}
    \ket{A_S(t)}\rangle&=\mathcal{S}\ket{A_S(0)}\rangle \nonumber \\
    &=\mathcal{S}\bigg|\sum_{ij}A_{ij}\ket{i}\bra{j}\bigg\rangle\bigg\rangle =\mathcal{S}\sum_{ij}A_{ij}\big|\ket{i}\bra{j}\big\rangle\big\rangle \nonumber \\
    &=\mathcal{S}\sum_{ij}A_{ij}\ket{ij}\rangle =\sum_{ij}A_{ij}\mathcal{S}\ket{ij}\rangle.
\end{align}
We change the indices of the vector $\ket{ij}\rangle \to \ket{k}$ for $k = 0, 1, 2, 3$ as $i, j\in\{0, 1\}$, such that $\ket{00}\rangle\to\ket{0}, \ket{01}\rangle\to\ket{1}, \ket{10}\rangle\to\ket{2}$, and $\ket{11}\rangle\to\ket{3}$. Using this, we can rewrite the above equation as
\begin{align}
    \ket{A_S(t)}\rangle&=\sum_{k}A_{k}\mathcal{S}\ket{k} \nonumber \\
    &=\sum_{k}A_{k}\times(\text{$k^{th}$ column of $\mathcal{S}$}).
\end{align}
On comparing this with Eq.~\eqref{rhoStEdag}, we obtain
\begin{align}
     \big|\Edag(\ket{i}\bra{j})\big\rangle\big\rangle\equiv\text{$k^{th}$ column of $\mathcal{S}$},
\end{align}
The knowledge of the adjoint map $\Edag$ is provided by the Hamiltonian of the total evolution. By calculating $\ket{\Edag(\ket{i}\bra{j})}$ we can obtain the superoperator $\mathcal{S}$ column by column.
Using $\mathcal{S}$, the bipartite OTOC in Eq.~\eqref{Openbotoc} is obtained.

\section{The swap operators}
The derivation of $G(t)$ makes use of the following identity involving the Haar integral over random unitary matrices 
\begin{align}
    \left( \int_ {U{(d)}}  U \otimes U^{\dagger}  d\mu(U) \right) = \frac{S}{d}.
\end{align}
The swap operators $S$ are introduced below. $U$ denotes a random unitary matrix. To illustrate how the swap operators are structured, a replica of the original Hilbert space is taken, such that $\mathcal{H}^{\prime}=\mathcal{H}_{A^{\prime}} \otimes \mathcal{H}_{B^{\prime}}$ having the same dimension as $\mathcal{H}$, i.e., $d={\rm dim}(\mathcal{H})={\rm dim}(\mathcal{H}^\prime)=d_Ad_B$.
Defining $S_{AA^{\prime}}$, that is supported over $\mathcal{H} \otimes \mathcal{H^{\prime}}$ and swaps only the indices of the $\mathcal{A}$ subsystem, we obtain the bipartite OTOC as
\begin{equation}
    G(t)=1-\frac{1}{d^2}{\rm Tr}(S_{AA^{\prime}}U_t^{\otimes2}S_{AA^{\prime}}U_t^{\dagger \otimes2}).
\end{equation}
The same expression is valid for $S_{BB^{\prime}}$ as well.
The structure of $S$, $S_{AA^{\prime}}$ and $S_{BB^{\prime}}$ are
\begin{align}
    S=\sum_{a,b,a^{\prime},b^{\prime}}\ket{aba^{\prime}b^{\prime}}\bra{a^{\prime} b^{\prime} a b }, \\
    S_{AA^{\prime}}= \sum_{a,b,a^{\prime},b^{\prime}}\ket{aba^{\prime}b^{\prime}}\bra{a^{\prime} b a b^{\prime} }, \\S_{BB^{\prime}}=\sum_{a,b,a^{\prime},b^{\prime}}\ket{aba^{\prime}b^{\prime}}\bra{a b^{\prime} a^{\prime} b }.
\end{align}
All of the above swap operators belong to $\mathcal{H}\otimes \mathcal{H^{\prime}}=\mathcal{H}_A \otimes \mathcal{H}_B \otimes 
  \mathcal{H}_A ^{\prime} \otimes \mathcal{H}_B^{\prime} $ Hilbert space. 

\bibliographystyle{apsrev4-2}
\bibliography{Bibilo}  

\end{document}